\documentclass[letter]{aa}
\usepackage{amssymb}
\usepackage{amsmath}
\usepackage{wasysym} 
\usepackage{txfonts}
\usepackage{balance}
\usepackage{natbib}
\bibpunct{(}{)}{;}{a}{}{,} 
\usepackage{color}
\usepackage{xspace}
\usepackage{bbm}
\usepackage{cancel}
\usepackage{graphicx}
\usepackage{ifthen}
\usepackage{float}
\newcommand{\e}[1]{\ensuremath{\times 10^{#1}}}

\newcommand{\rhos}{\ensuremath{\rho_\text{s}}\xspace}
\newcommand{\amax}{\ensuremath{a_\text{max}}\xspace}

\newcommand{\uf}{\ensuremath{u_\text{f}}\xspace}
\newcommand{\cs}{\ensuremath{c_\mathrm{s}}\xspace}

\newcommand{\Siggas}{\ensuremath{\Sigma_\mathrm{g}}\xspace}

\newcommand{\Fmm}{\ensuremath{{F_\text{1mm}}}\xspace}
\newcommand{\alphamm}{\ensuremath{\alpha_\text{1-3mm}}\xspace}
\newcommand{\betamm}{\ensuremath{\beta_\text{1-3mm}}\xspace}
\newcommand{\alphat}{\ensuremath{\alpha_\text{t}}\xspace}
\definecolor{mygray}{gray}{.5}
\newcommand{\drop}[1]{\xspace}
\makeatletter
\if@referee
    \newcommand{\change}[2]{\textcolor{red}{#1\xspace}}

    \usepackage[nolists,noheads]{endfloat}
\else

   \newcommand{\change}[2]{\textcolor{black}{#1\xspace}}         

\fi
\makeatother
\usepackage[%
pdfauthor={Tilman Birnstiel},
pdftitle={Testing the theory of grain growth and fragmentation by millimeter observations of protoplanetary disks},
pdfstartview=FitH,
linkcolor=blue,
anchorcolor=blue,
citecolor=blue,
filecolor=blue,
menucolor=blue,
urlcolor=blue,
colorlinks=true,
linktocpage=true]{hyperref}

\defcitealias{Ricci:2010p9423}{R10}

\begin{document}
\title{Testing the theory of grain growth and fragmentation by millimeter observations of protoplanetary disks}
\titlerunning{Testing the theory of grain growth by millimeter observations}
\author{
  T.~Birnstiel\inst{1} \and
  L.~Ricci\inst{2} \and
  F.~Trotta\inst{2}\fnmsep\inst{3} \and
  C.P.~Dullemond\inst{1} \and
  A.~Natta\inst{3} \and
  L.~Testi\inst{2}\fnmsep\inst{3} \and
  C.~Dominik\inst{4}\fnmsep\inst{5} \and
  T.~Henning\inst{1} \and
  C.W.~Ormel\inst{1} \and
  A.~Zsom\inst{1}
  }
\authorrunning{T.~Birnstiel~et~al.}
\institute{
    Max-Planck-Institut f\"ur Astronomie, K\"onigstuhl 17, 69117 Heidelberg, Germany
    \and
    European Southern Observatory, Karl-Schwarzschild-Strasse 2, 85748 Garching, Germany
    \and
    Osservatorio Astrofsico di Arcetri, INAF, Largo E. Fermi 5, 50125 Firenze, Italy
    \and
    Astronomical Institute ``Anton Pannekoek'', University of Amsterdam, PO Box 94249, 1090 GE Amsterdam, The Netherlands
    \and
    Afdeling Sterrenkunde, Radboud Universiteit Nijmegen, Postbus 9010, 6500 GL Nijmegen, The Netherlands
}
\date{\today}
\abstract
{Observations at sub-millimeter and mm wavelengths will in the near future be able to resolve the radial dependence of the mm spectral slope in circumstellar disks with a resolution of around a few~AU at the distance of the closest star-forming regions.}
{We aim to constrain physical models of grain growth and fragmentation by a large sample of (sub-)mm observations of disks around pre-main sequence stars in the Taurus-Auriga and Ophiuchus star-forming regions.}
{State-of-the-art coagulation/fragmentation and disk-structure codes are coupled to produce steady-state grain size distributions and to predict the spectral slopes at (sub-)mm wavelengths.}
{This work presents the first calculations predicting the  mm spectral slope based on a physical model of grain growth. Our models can quite naturally reproduce the observed mm-slopes, but a simultaneous match to the observed range of flux levels can only be reached by a reduction of the dust mass by a factor of a few up to about 30 while keeping the gas mass of the disk the same. This dust reduction can either be due to radial drift at a reduced rate or during an earlier evolutionary time (otherwise the predicted fluxes would become too low) or due to efficient conversion of dust into larger, unseen bodies.}
{}
\keywords{accretion, accretion disks -- circumstellar matter -- stars: formation, pre-main-sequence -- infrared: stars}
\maketitle
\section{Introduction}                  \label{sec:intro}
Circumstellar disks play a fundamental role in the formation of stars as most of the stellar material is believed to be transported through the disk before being accreted onto the star \citep{LyndenBell:1974p1945}. At the same time circumstellar disks are thought to be the birth places of planets. Understanding the physics of circumstellar disks is therefore the key to some of the most active fields of astrophysical research today.

However, observing these disks in order to learn about the physical processes taking place in their interior is a challenging task.
\citet{Strom:1989p9475} and \citet{Beckwith:1990p3768} were the first to use observations at mm-wavelengths to confirm that many of the observed pre-main sequence (PMS) stars showed excess radiation above the spectrum of a T Tauri star.
While these single dish observations provided valuable insight in dust masses (since mm observations probe not only the thin surface layers, but the bulk of the dust mass in the mid-plane), recent developments in the field of \mbox{(sub-)mm} interferometry grant the possibility to constrain models of disk structure and evolution of protoplanetary disks by fitting parametric models to the observed radial profiles \citep[e.g.,][]{Andrews:2009p7729,Isella:2009p7470}. Spatially resolving the disks is important since it ensures that low millimeter spectral slopes are not just an artifact of high optical depth.

Today, mm spectral slopes are known for quite a number of disks and spatially resolved observations indicate that the low values measured in these samples are related to grain growth \citep[e.g.,][]{Testi:2003p3390,Natta:2004p3169,Rodmann:2006p8905}. Grains are believed to collide and stick together by van der Waals forces, thus forming larger and larger aggregates \citep{Dominik:1997p9440,Poppe:2000p9447,Blum:2008p1920}. Due to this loose binding, collisions with velocities in excess of a few m~s$^{-1}$ may lead to fragmentation of the aggregates.

Larger samples of radially resolved mm spectral slopes are expected in the near future, but still no study so far interpreted mm observations using simulated grain size distributions but rather used simple parametric power-law distributions with an upper size cut-off. In this work, we use a state of the art dust grain evolution code \citep[similar to][]{Brauer:2008p215,Birnstiel:2010p9709} to derive steady-state grain distributions where grain growth and fragmentation effects balance each other. We self-consistently solve for the grain size distributions and the disk structure to predict fluxes at mm wavelengths and the radial dependence of the mm spectral index. Comparing these results to observed values in the Taurus and Ophiuchus star-forming regions allows us to test predictions of the theory of grain growth/fragmentation and to infer constraints on grain properties such as the critical collision velocity and the distribution of fragments produced in collision events.

Grains orbiting at the Keplerian velocity in a laminar gas disk feel a constant head wind (caused by the gas rotating slightly sub-keplerian) which forces them to spiral inwards \citep{Weidenschilling:1977p865}. If this drag is as efficient as laminar theory predicts, all dust particles which are necessary to explain the observed spectral indices, would quickly be removed \citep[see][]{Brauer:2007p232}. In this Letter, we therefore assume that radial drift is halted by an unknown mechanism. Under this assumption, we find that low values of the mm spectral index can be explained by the theory. We show that in order to explain the observed flux levels, the amount of observable\footnote{by ``visible'' or ``observable'' dust we mean the dust particles which are responsible for most of the thermal continuum emission at \mbox{(sub-)mm} wavelengths which are typically smaller than a few centimeter in radius.} dust needs to be reduced by either reducing the dust-to-gas ratio (perhaps by radial drift at a intermediate efficiency or during an earlier evolutionary epoch) or by dust particle growth beyond centimeter sizes. 

\section{Model description}             \label{sec:model}
\subsection{Disk model}                 \label{sec:model_disk}
We consider disks around a PMS star with a mass of 0.5~$M_{\odot}$, bolometric luminosity of 0.9~$L_{\odot}$ and effective temperature of 4000~K, at a distance of 140~pc, which are typical values for the sample of low-mass PMS stars studied in \citet[][hereafter \citetalias{Ricci:2010p9423}]{Ricci:2010p9423}. To derive the disk structure we adopted a modified version of the two-layer models of passively irradiated flared disks developed by \citet{Dullemond:2001p9307} (following the schematization by \citealp{Chiang:1997p1986}), in which we have relaxed the common assumption that dust grain properties are constant throughout the disk. For the disk surface density we adopted the self-similar solution for a viscous disk \citep[see][]{LyndenBell:1974p1945} with parameters lying in the ranges observationally constrained by \citet{Andrews:2009p7729}. The surface density gradient $\gamma$ and the characteristic radius $R_\mathrm{c}$ \citep[for the definitions, see][]{Hartmann:1998p664} are assumed to be $\gamma=1$ and $R_\mathrm{c}=60$~AU, respectively. Throughout this work we assume a constant dust-to-gas mass ratio of 1\%.

\begin{table}
\caption{Parameters of the model grid: $M_\text{disk}$ is the total disk mass, $\alphat$ is the turbulence parameter, \uf is the critical collision velocity, $f_\text{vac}$ is the grain volume fraction of vacuum and $\xi$ is the index of the distribution of fragments (see Eq.~\ref{eq:n_frag}). The parameters of the fiducial model are highlighted in bold face.}
\label{tab:model_grid}
\centering
\begin{tabular}{lr|llll}
\hline\hline
parameter & \multicolumn{4}{c}{values}\\
\hline\\[-0.25cm]
    $M_\mathrm{disk}$ &$[M_\odot]$ & 5\e{-3}                 & $\mathbf{1\e{-2}}$          & 5\e{-2} & 1\e{-1}\\
    $\alphat$         &            & $\mathbf{5\e{-4}}$      & 1\e{-3}                     & 5\e{-3}                    & -      \\
    \uf               &[m/s]       & 1                       & \textbf{3}                  & 10                         & -      \\
    $f_\text{vac}$    &[\% vol.]   & \textbf{10}             & 30                          & 50                         & -      \\
    $\xi$             &            & 1.0                     & \textbf{1.5}                & 1.8                        & -      \\
\hline
\end{tabular}
\end{table}

\subsection{Dust model}                 \label{sec:model_dust}
We use a coagulation/fragmentation code as described in \citet{Brauer:2008p215} and \citet{Birnstiel:2010p9709} to simulate the growth of dust particles. Particles grow due to mutual collisions (induced by Brownian motion and by turbulence, see \citealp{Ormel:2007p801}) and subsequent sticking by van der Waals forces. We assume the dust particles to be spheres of internal density \rhos and vary \rhos to account for porosity effects. However, we do not treat a dynamic porosity model \citep[see][]{Ormel:2007p7127,Zsom:2008p7126}.

With increasing collision velocity $\Delta u$, the probability of sticking decreases and fragmentation events start to become important. In this Letter, we use the fragmentation probability  
\begin{equation}
p_\text{f} = \left\{
\begin{array}{ll}
0&                              \text{if } \Delta u < \uf - \delta u\\
1&                              \text{if } \Delta u > \uf\\
1-\frac{\uf-\Delta u}{\delta u}&    \text{else}
\end{array}
\right.
\end{equation}
where \uf is the collision velocity above which particles are assumed to fragment and $\delta u$ is the transition width between coagulation and fragmentation (taken to be 0.2 \uf). Recent studies of collision experiments \citep{Guttler:2010p9745} and numerical simulations \citep{Zsom:2010p9746} indicate that there is also a regime in which particles may bounce. However, since this topic is still not well understood, we will omit these effects in this work. 

Radial drift is an, as yet, unsolved problem \citep{Brauer:2007p232,Birnstiel:2009p7135}. However there are several effects such as spiral wave structure \citep[e.g.,][]{Cossins:2009p3794}, density sinks \citep[e.g.,][]{Brauer:2008p212} or zonal flows \citep[e.g.,][]{Johansen:2009p7441} which may reduce the effectiveness of radial drift. In this paper we assume that radial drift is ineffective\drop{, unlike predicted for laminar disks \citep[see][]{Whipple:1972p4621,Weidenschilling:1977p865}} since we focus on the question whether observations can be explained through the physics of grain growth and fragmentation. The question to answer in this case is not, how to retain particles at these radii, but rather how to create them there in the first place.

To investigate this problem, we simulate the physics of particle growth and fragmentation until a steady state between both processes develops. Since the relative velocities for particles typically increase with grain radius, we can relate the fragmentation velocity to a certain grain size \citep[which defines the ``fragmentation barrier'', see][]{Birnstiel:2009p7135}
\begin{equation}
a_\mathrm{max} \simeq \frac{2\Siggas }{\pi \alphat \rho_\mathrm{s}} \cdot \frac{\uf^2}{\cs^2},
\label{eq:a_max}
\end{equation}
above which particles fragment (with \Siggas, $\alphat$ and \cs being the gas surface density, the turbulence parameter and the sound speed, respectively).
\uf and \alphat are assumed to be radially constant with \alphat values within a range expected from theoretical (see \citealp{Johansen:2005p8425}; Dzyurkevich et al., 2010, in press) and observational works \citep[see][]{Andrews:2009p7729}.
Grains which reach \amax will experience high velocity collisions causing them to be eroded or even completely fragmented. The resulting fragments can again contribute to growth processes at smaller sizes and the grain size distribution will at some point reach a steady state where gain and loss terms caused by coagulation and by fragmentation cancel out at all sizes.

Particles will need a certain time to grow to the fragmentation barrier. The time to reach the steady state will therefore be a few of these growth time scales. Depending on the distance to the central star, the steady state is typically reached after a few thousand years at 1~AU up to about 1~Myr at 100~AU. The mean ages of the sources in our sample are $\approx 2$~Myr and $\approx 0.5-1$~Myr for the Taurus and Ophiuchus PMS stars, respectively. Since radii around $40-80$~AU dominate the observed emission at \mbox{(sub-)mm} wavelengths, we expect most of the samples to be in or at least close to a steady state. 

If the highest collision velocity that turbulent motion induces (depending on $\alphat$ and \cs) is smaller than the critical collision velocity \uf, then (at least some) particles do not fragment (i.e. the break through the fragmentation barrier) and a steady state is never reached. Due to this scenario, some of the possible combinations of the parameter values (see Table~\ref{tab:model_grid}) do not reach a steady state and are therefore not included in the results. 

The shape of the steady state grain size distributions is influenced mainly by five parameters: the previously mentioned $\alphat$, \uf, \Siggas, the temperature $T$ (through the sound speed \cs), and by the prescription of fragmentation. In our models, we assume the distribution of fragments to follow a power-law number density distribution,
\begin{equation}
n(m) \propto m^{-\xi},
\label{eq:n_frag}
\end{equation} 
with an upper end at $m_\mathrm{f}$. We consider fragmentation and cratering, as described in \citet{Birnstiel:2010p9709}.
Recent experiments suggest $\xi$ values between 1.07 and 1.37 \citep[see][]{Guttler:2010p9745}. In this work, we consider $\xi$ values between $1.0$ and $1.8$.

To calculate the dust opacity of a given grain size distribution we adopted the same dust grain model taken in \citetalias{Ricci:2010p9423}, i.e. porous composite spherical grains made of astronomical silicates, carbonaceous materials and water ices (see \citetalias{Ricci:2010p9423} for the references to the optical constants). The ratio between the fractional abundance of each species comes from \citealp{Semenov:2003p9622} and models with three different porosities have been considered in this Letter (see Table~\ref{tab:model_grid}).
We used the Bruggeman mixing theory to combine the refractive indices of the different materials and to calculate the dust opacity of the composite grains.
The opacity induces probably the largest uncertainties in our calculations as grain composition, grain structure and temperature effects may lead to largely different opacities \citep[see, for example][]{Henning:1996p4181}.

\subsection{Comparison to observations} \label{sec:model_comparison}
We compare the \mbox{(sub-)mm} SED generated by our models with observational data of \citetalias{Ricci:2010p9423} and Ricci et al. (in prep), more specifically the flux at 1~mm (\Fmm) and the spectral index between 1 and 3~mm ($F(\lambda) \propto \lambda^{-\alphamm}$). The samples considered include all the class II disks in the Taurus-Auriga and $\rho$-Oph star-forming regions respectively for which both the central PMS star and the disk are observationally well characterized through optical-NIR spectroscopy/photometry and \mbox{(sub-)mm} photometry/interferometry. To calculate the dust opacity as a function of wavelengths and radius, and the temperature in the disk mid-plane, we iterated the two-layer disk model (keeping the profile of $\Siggas$ constant in time) with the dust model described above until convergence is reached. Once the physical structure of the disk is determined, the two-layer disk models return the disk SED which can be compared to the observations.

The influence of the different parameters on the calculated \alphamm values can mostly be understood by a simple model for a dust distribution, as used in \citetalias{Ricci:2010p9423} (cf. Fig.~3 in \citetalias{Ricci:2010p9423}):
for maximum grain sizes much smaller than the observed wavelengths, the spectral index of the dust opacity \betamm ($\kappa(\lambda) \propto \lambda^{-\betamm}$) is constant, while it decreases for \amax values larger than sub-mm sizes. In between (at a few tenth of a mm), there is a peak which is caused by an increased opacity of grains with sizes similar to the observed wavelength. The relation between $\alphamm$ and $\betamm$ depends on the emitting spectrum and the optical depth. For a completely optically thin disk in the Rayleigh-Jeans regime $\betamm=\alphamm-2$. However, if the emitted spectrum deviates from the Rayleigh-Jeans limit, then $\betamm \gtrsim \alphamm-2$. In our models, $\alphamm-\betamm$ turns out to be typically between $1.4$ and $1.7$ if \amax is outside of the peak of opacity.

\section{Results}                       \label{sec:results}
\subsection{Sub-mm fluxes and spectral indices}                 \label{sec:results_fluxes}
For all possible combinations of the parameters shown in Table~\ref{tab:model_grid}, we solved for the steady-state grain size distributions and derived the \alphamm and \Fmm values. As noted before, some of the models do not result in a steady state and are, therefore, not shown here.

\begin{figure}[t]
  \centering
  \resizebox{0.9\hsize}{!}{\includegraphics{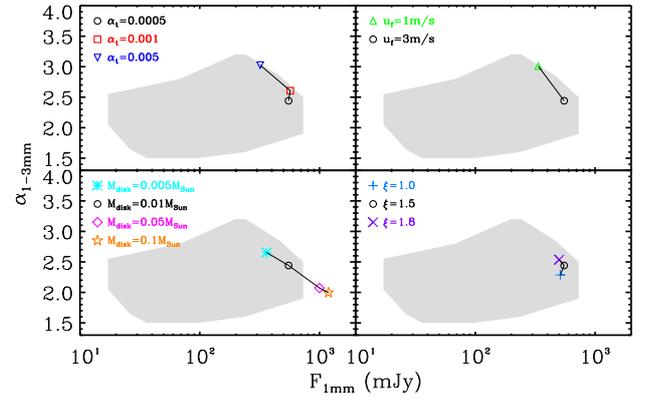}}
  \caption{Influence of the parameters \alphat (top left), fragmentation velocity (top right), disk mass (bottom left) and grain porosity (bottom right) on the observed fluxes and spectral indices. The black circle denotes the fiducial model whose parameters are given in Table~\ref{tab:model_grid}. The grey area represents the region in which the observed sources lie (see Fig.~\ref{fig:clouds}).}
  \label{fig:multiplot}
\end{figure}

The top left panel of Fig.~\ref{fig:multiplot} shows the influence of the turbulence parameter \alphat. According to Eq.~\ref{eq:a_max}, the maximum grain size increases if \alphat decreases. Depending on where \amax lies with respect to the opacity peak (see \citetalias{Ricci:2010p9423}, Fig.~3), \alphamm can increase or decrease with increasing \alphat. In the simulations presented here, \amax is typically so large that increaseing \alphat predicts larger spectral slopes.

\amax is more sensitive to \uf (cf. top right panel in Fig.~\ref{fig:multiplot}): the maximum grain size $a_\text{max}$ is proportional to $\uf^2$, therefore a change of \uf by a factor of about 3 significantly changes \alphamm by increasing the grain size by about one order of magnitude.
However many models with a fragmentation velocity of 10~m/s never reach a steady state. It is therefore not possible to explain lower \alphamm values by a further increase of \uf alone.

The way that $M_\text{disk}$ influences \Fmm and \alphamm is twofold. Firstly, a decrease in $M_\text{disk}$ (assuming a constant dust-to-gas ratio and a fixed shape of the disk \change{surface density}, i.e. not varying $R_\mathrm{c}$ and $\gamma$, see Section~\ref{sec:model_disk}), reduces the amount of emitting dust and thus \Fmm. Secondly, such a reduction in gas mass also reduces \amax (Eq.~\ref{eq:a_max}), which tends to increase \alphamm. This combined trend is seen in Fig.~\ref{fig:multiplot}. Hence, in order to explain faint sources with low \alphamm, the amount of emitting dust has to be reduced while the disk gas mass stays large. This effect could be achieved in two ways: the amount of dust could be reduced by radial drift at a reduced rate (full radial drift would quickly remove all mm-sized grains, see \citealp{Brauer:2007p232}) or only the ``visible'' amount of dust is reduced if some of the dust is already contained in larger bodies. \change{This latter case is predicted
by our non-steady-state distribution models and will be discussed in more details in a forthcoming
paper.}{}

\begin{figure}[b!t]
  \centering
  \resizebox{0.9\hsize}{!}{\includegraphics{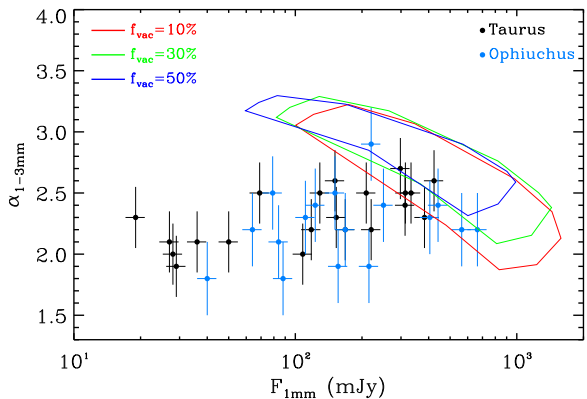}}
  \caption{Observed fluxes at mm-wavelengths of the Taurus (black dots) and the Ophiuchus (blue dots) star-forming regions (see \citetalias{Ricci:2010p9423} and Ricci et al. 2010, in prep) and the areas covered by the simulation results for different vacuum fractions of the grains (varying all other parameters according to Table~\ref{tab:model_grid}).}
  \label{fig:clouds}
\end{figure}

In general, lower values of $\xi$ translate to shallower grain-size distribution, which result in lower values of \betamm \citep[see][]{Draine:2006p9433}.
The lower right panel in Fig.~\ref{fig:multiplot} does not seem to indicate a strong dependence on $\xi$, however lower values of $\xi$ (around ~ 1) seem to be closer to the observations especially at large fluxes.

Figure~\ref{fig:clouds} shows the areas which are covered by our sets of simulations for different porosities in comparison to the observational samples. It can be seen that only the brightest sources are covered by the simulations. The trend of larger \alphamm for larger vacuum fraction seems to be in contradiction with Eq.~\ref{eq:a_max}, since smaller grain volume density leads to larger \amax. However in this case, the opacity is much more affected by changing the grain structure: reducing the grains vacuum fraction increases the spectral index at sub-mm while it is reduced for longer wavelength. Therefore opacity effects outweigh the smaller changes in \amax.
A more thorough analysis of opacity effects is beyond the scope of this letter, however it seems implausible that the large spread in the observations can be explained by different kinds of grains alone \citep[see][]{Draine:2006p9433}.

\subsection{Radial profiles of the dust opacity index}                 \label{sec:results_beta}
\begin{figure}[bt]
  \centering
\resizebox{0.9\hsize}{!}{\includegraphics{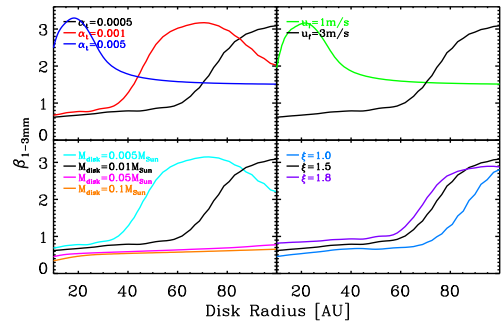}}
  \caption{Predicted profiles of the dust opacity index at mm-wavelengths for different variations of the fiducial model. The colors correspond to the parameters shown in Fig.~\ref{fig:multiplot}.}
  \label{fig:betamultiplot}
\end{figure}

The presented models also compute \alphamm as function of radius. From the point of view of comparison to observations, this is somewhat premature since observational methods are not yet able to provide reliable radial profiles of \alphamm (e.g., \citealp{Isella:2010p9438}, Banzatti et al., in prep.). 
However, the predicted radial dependence of \betamm (shown in Fig.~\ref{fig:betamultiplot}) agrees with the observations so far. It can be seen that the shape of most models looks similar, slightly increasing from \betamm-values around 0.5 at 10~AU up to around 1.5 at 100~AU. The reason for this is that \amax depends on the ratio of surface density over temperature. Under typical assumptions, \amax will decrease with radius. An upper grain size which is decreasing with radius and stays outside the peak in the opacity results in \betamm increasing with radius (cf. Fig.~3 in \citetalias{Ricci:2010p9423}). If the radially decreasing upper grain size \amax reaches sizes just below mm, then the peak in opacity will produce also a peak in the radial profile of \betamm (the size of which depends much on the assumed opacity), which can be seen in Fig.~\ref{fig:betamultiplot}. Thus, even though \amax is monotone in radius, \betamm does not need to be monotone. 

\section{Discussion and Conclusions}                   \label{sec:conclusions}
In this Letter, we present the first in-depth comparison of simulated grain size distributions and observed mm spectral indices of YSOs in the Taurus and the Ophiuchus star-forming regions. Additionally we present the first predictions of the radial profile of the dust opacity index at mm wavelength which are consistent with the limits set by \citet{Isella:2010p9438}.

Low values of the observed mm-slopes are quite naturally reproduced by our models, favoring low values of $\xi$ and \alphat as well as fragmentation threshold velocities above 1~m~s$^{-1}$. However, a simultaneous match to the observed range of flux levels requires a reduction of the dust mass by a factor of a few up to about 30. This over-prediction of fluxes cannot be fixed by simply reducing the disk mass since the predicted \alphamm would be too large for smaller disk masses.
Opacities induce a large uncertainty in the flux levels. However, considering the results of \citet{Draine:2006p9433}, it seems implausible that the large spread in observed fluxes for different disks with similar \alphamm (which is probably even larger as very faint disks are not contained in the sample) can be explained by different grain mineralogy alone.

The aforementioned reduction of observable dust could be due to radial drift at a reduced rate or during an earlier epoch (drift has been artificially suppressed in this work in order to explain the low values of \alphamm by $\gtrsim 1$~mm sized grains). Another possible explanation is grain growth to even larger sizes as these bodies have a small opacity coefficient per unit mass.

\change{Finally, a different dependence between \alphamm and the observed flux \Fmm might also originate from disk surface densities profiles that differ from what we have assumed in this work. This possibility, as well as the effect of a different dust composition, will be considered in a future work.}{Future work will also consider different opacity and composition of the grains as well as time dependent simulations of the gas surface density and the dust evolution.}
\begin{acknowledgements}
We like to thank Henrik Beuther and J\"urgen Blum for useful discussions. L. R. and A. Z. acknowledge support of the International Max-Planck-Research School. C.W.O. is supported by a grant from the Alexander von Humboldt foundation. We also thank Andrea Isella for a helpful referee report.
\balance
\end{acknowledgements}
\bibliographystyle{aa}
\bibliography{letter-2010-01}
\end{document}